\documentclass[onecolumn,journal]{IEEEtran}
\usepackage{amsmath,amsfonts}
\usepackage{algorithmic}
\usepackage{algorithm}
\usepackage{array}
\usepackage[caption=false,font=normalsize,labelfont=sf,textfont=sf]{subfig}
\usepackage{textcomp}
\usepackage{xcolor}
\usepackage{stfloats}
\usepackage{url}
\usepackage{verbatim}
\usepackage{graphicx}
\usepackage{cite}
\def\BibTeX{{\rm B\kern-.05em{\sc i\kern-.025em b}\kern-.08em
		T\kern-.1667em\lower.7ex\hbox{E}\kern-.125emX}}
\newtheorem{thm}{Theorem}

\newtheorem{cor}{Corollary}

\newtheorem{defn}{Definition}

\newtheorem{rem}{Remark}
\newcommand{\squeezeequ}{\medmuskip=2mu \thinmuskip=1mu \thickmuskip=3mu}

\newcommand{\Dsupersqueezeequ}{\medmuskip=0.5mu \thinmuskip=0mu \thickmuskip=1mu \nulldelimiterspace=-1pt \scriptspace=0pt}
\newcommand{\Tsupersqueezeequ}{\medmuskip=0.1mu \thinmuskip=0mu \thickmuskip=0.1mu \nulldelimiterspace=-1pt \scriptspace=0pt}

\begin{document}

\title{Sparse Regression Codes for Secret Key Agreement: Achieving Strong Secrecy and Near-Optimal Rates for Gaussian Sources}

\author{\IEEEauthorblockN{Emmanouil M. Athanasakos and Hariprasad Manjunath\\}

			\thanks{E.M. Athanasakos is with the National and Kapodistrian University of Athens, Dpt. of Informatics and Telecommunications, Greece, e-mail: emathan@di.uoa.gr. \\
			H.Manjunath is with the Chanakya University, Bengaluru, India, e-mail: mhariprasadkansur@gmail.com}%
}

\markboth{PREPRINT}%
{Athanasakos and Manjunath: Sparse Regression Codes for Secret Key Agreement: Achieving Strong Secrecy and Near-Optimal Rates for Gaussian Sources}


\maketitle

\begin{abstract}
Secret key agreement from correlated physical layer observations is a cornerstone of information-theoretic security. This paper proposes and rigorously analyzes a complete, constructive protocol for secret key agreement from Gaussian sources using Sparse Regression Codes (SPARCs). Our protocol systematically leverages the known optimality of SPARCs for both rate-distortion and Wyner-Ziv (WZ) coding, facilitated by their inherent nested structure. The primary contribution of this work is a comprehensive end-to-end analysis demonstrating that the proposed scheme achieves near-optimal secret key rates with strong secrecy guarantees, as quantified by a vanishing variational distance. We explicitly characterize the gap to the optimal rate, revealing a fundamental trade-off between the key rate and the required public communication overhead, which is governed by a tunable quantization parameter. Furthermore, we uncover a non-trivial constrained optimization for this parameter, showing that practical constraints on the SPARC code parameters induce a peak in the achievable secret key rate. This work establishes SPARCs as a viable and theoretically sound framework for secure key generation, providing a compelling low-complexity alternative to existing schemes and offering new insights into the practical design of such protocols.
\end{abstract}


\section{Introduction}
\IEEEPARstart{I}{nformation}-theoretic security leverages physical layer properties to establish secure communication links, offering provable security guarantees based on fundamental physical limits. Secret key (SK) agreement, a core primitive alongside the wiretap channel \cite{bloch_book}, focuses on extracting a shared secret key from correlated random observations available to legitimate parties. This is particularly relevant in communication scenarios where dedicated secure channels are unavailable or impractical, such as securing wireless sensor networks, enabling device pairing in IoT ecosystems based on shared environmental noise, or establishing secure links in vehicular networks using correlated fading measurements. The SK agreement process is typically assisted by communication over an insecure public channel, which is also accessible to potential eavesdroppers. Groundbreaking work by Maurer \cite{Maurer_93} and Ahlswede and Csiszár \cite{Ahlswede_93} first demonstrated how these noisy correlations could be distilled into secret keys. This foundational framework was subsequently extended to continuous sources, with significant focus on Gaussian sources \cite{Nitinawarat_12, Watanabe_10}, providing precise characterizations of the optimal secret key rate achievable under public communication constraints. The overall SK generation process generally involves two critical phases: information reconciliation, where legitimate parties use the public channel to agree on a common sequence from their correlated observations, and privacy amplification, where they distill a highly secure key from this common sequence, ensuring it is statistically independent of the eavesdropper's knowledge \cite{bloch_book}.

While the theoretical limits of SK agreement, representing the ultimate performance benchmark, are well understood, the development of practical, low-complexity coding schemes that approach these limits remains a pivotal and active research challenge. Significant progress has been made by leveraging structured codes. Polar codes, for instance, have been shown to be capacity-achieving for SK agreement in various source and channel models \cite{Chou_15, Renes_SKA}, offering elegant constructions with efficient encoding and decoding algorithms. Independently, nested lattice codes were developed and proven effective \cite{Ling_13, Vadenka}, achieving strong secrecy and near-optimal key rates, often up to a small constant gap, with polynomial complexity. These contributions represent powerful solutions. However, the exploration of alternative coding frameworks is crucial as they can offer different performance-complexity trade-offs, distinct structural properties beneficial for advanced multi-user scenarios, or better suitability for specific source statistics or hardware platforms. This motivates our investigation into Sparse Regression Codes (SPARCs).

SPARCs, introduced by Barron and Joseph \cite{barron_1,barron_2}, represent a capacity-achieving class of codes for the Additive White Gaussian Noise (AWGN) channel, built upon principles of sparse linear regression. Their inherent characteristics make them compelling candidates for constructing SK agreement protocols. Firstly, SPARCs are proven to achieve the optimal rate-distortion function for Gaussian sources \cite{venka_12_lossy, venk_12_isit}. This is directly applicable to the initial source quantization step common in many SK protocols, ensuring maximal information retention for a given compression. Secondly, the SPARC framework elegantly supports binning through a nested code structure \cite{venk_12_aller}. This binning capability is fundamental for implementing efficient information reconciliation schemes, particularly those based on Wyner-Ziv (WZ) coding. Thirdly, SPARCs have demonstrated optimality for the WZ lossy source coding problem itself \cite{venk_12_aller}, which precisely models the reconciliation task where one party (e.g., Bob) decodes the source sequence observed by another (Alice) using its own correlated observation as side information. Furthermore, our prior work has shown the utility of SPARCs in achieving secrecy capacity for the Gaussian wiretap channel \cite{Athan_20}, underscoring their potential in security-centric applications.

In this paper, we harness these advantageous features of SPARCs to design and rigorously analyze a complete SK agreement protocol tailored for correlated Gaussian sources in the presence of an eavesdropper. Our protocol explicitly employs SPARCs for both the initial quantization of Alice's source observation by leveraging their rate-distortion optimality and the subsequent information reconciliation phase capitalizing on their WZ optimality and nested structure. Privacy amplification is then performed using standard universal hash functions to extract the final secret key. We conduct a thorough analysis, adhering to the established steps of reconciliation and privacy amplification, carefully adapted for the specifics of the SPARC construction. Our main theoretical result demonstrates that this SPARC-based protocol achieves strong secrecy, ensuring the generated key is statistically independent of the eavesdropper's information including all public messages. Moreover, the protocol attains secret key rates that closely approach the fundamental upper bound established in \cite{Nitinawarat_12, Watanabe_10}. The gap to this optimal rate is explicitly characterized, highlighting the trade-offs involved. We uncover a non-trivial constrained optimization problem for the practical design of the protocol. Our numerical analysis shows that constraints on the SPARC code parameters create a peak in the achievable key rate. We formalize this optimization problem, showing that the optimal operating point represents the best performance achievable for a given SPARC family, an insight crucial for practical implementations. This work, therefore, establishes SPARCs as a theoretically sound and practically promising alternative coding framework for secret key generation from Gaussian sources, complementing existing approaches.

The remainder of this paper is organized as follows: Section II provides a detailed review of the pertinent background on SPARCs, emphasizing their nested structure and WZ optimality. Section III introduces the system model for SK generation from correlated Gaussian sources and outlines the proposed SPARC-based protocol steps. Section IV presents our main results, including the achievable rates, strong secrecy guarantees, and an analysis of the optimality gap. The numerical illustrations of the main theoretical results are shown in Section V. Finally, Section VI concludes the paper, summarizes the key findings, and discusses potential avenues for future research.

\section{Preliminaries and Notation}\label{sec_prelim}
This section reviews the fundamentals of SPARCs, focusing on their basic construction, their optimality for Gaussian source and channel coding, the crucial nested property that enables binning, and their application to achieving the WZ rate. These properties form the building blocks for the secret key agreement protocol developed in this paper.

\subsection{SPARCs Basics}
A SPARC is defined by a dictionary matrix $\mathbf{A}$ of dimensions $n \times N$, whose elements are typically drawn independently from $\mathcal{N}(0,\frac{1}{n})$, where $N = ML$. A codeword $\mathbf{x} \in \mathbb{R}^n$ is constructed as a linear combination of $L$ columns of $\mathbf{A}$, that is $	\mathbf{x} = \mathbf{A}\beta$, where $\beta$ is an $N$-length sparse coefficient vector. Specifically, $\beta$ is structured into $L$ sections, each of length $M$. Within each of the $L$ sections, exactly one coefficient is non-zero, and its value is typically set to a constant $c = \sqrt{P/\zeta}$, where $P$ is related to the codeword power and $\zeta$ depends on the specific application (e.g., $\zeta=L$ if each chosen column contributes on average $P/L$ to the total power, or $\zeta=1$ if $c$ is related to the amplitude of a single section's contribution). For consistency with later use (e.g., WZ where $U$ might have its own power constraint), we can assume $c$ is appropriately chosen.

The parameters $L$ and $M$ are chosen to satisfy certain relationships for achieving a target rate $R$ (in nats per source symbol). For instance, $M^L = e^{nR}$ (or $L \log M = nR$ for large $L$,$M$). Another common parameterization involves relating $M$ to $L$ via $M = L^\alpha$, where $\alpha > 0$ is termed the section size rate. This leads to $\alpha L \log L \approx nR$.

SPARCs were introduced as capacity-achieving codes for the AWGN channel \cite{barron_1}. A significant property for source coding applications is their ability to achieve the optimal rate-distortion function for Gaussian sources, as stated in the following Theorem.

\begin{thm}  \label{Thm_lossy_SPARCs}
	\cite{venka_12_lossy} Let $X^n$ be an i.i.d. sequence from a Gaussian source $X \sim \mathcal{N}(0, \sigma_X^2)$. For any target distortion $D\in (0,\sigma_X^2)$, SPARCs can achieve the optimal rate-distortion function $R^{\star}(D)=\frac{1}{2}\log \frac{\sigma_X^2}{D}$ with a probability of error
	\begin{align}
		P_e(\mathcal{C}_n,D) = \mathbb{P}\left(|X^n-h(g(X^n))|^2 > D\right) \label{prob_of_err}
	\end{align}
	that vanishes as $n \to \infty$. Here $g(\cdot)$ and $h(\cdot)$ are the minimum-distance encoder and decoder mappings, respectively. This optimality is achieved provided the code parameters $n, M, L, \alpha$ are chosen according to conditions specified in \cite{venka_12_lossy}, such as $\alpha > \frac{2.5R}{R-1+D/\sigma_X^2}$ 
\end{thm}

\begin{rem}
	Theorem 1 implies that for $ v^* < D/\sigma_X^2$, where $v^* \approx 0.2032$ is the solution to $1-v+\frac{1}{v}\log v = 0$ as defined in \cite{venka_12_lossy}, a larger section size rate $\alpha$ is needed to ensure SPARCs achieve the optimal rate-distortion performance across all distortions in $(0, \sigma_X^2)$.
\end{rem}

\subsection{Nested Property}
\begin{figure}[!htb]
	\centering
	\def\svgwidth{0.7\columnwidth}
	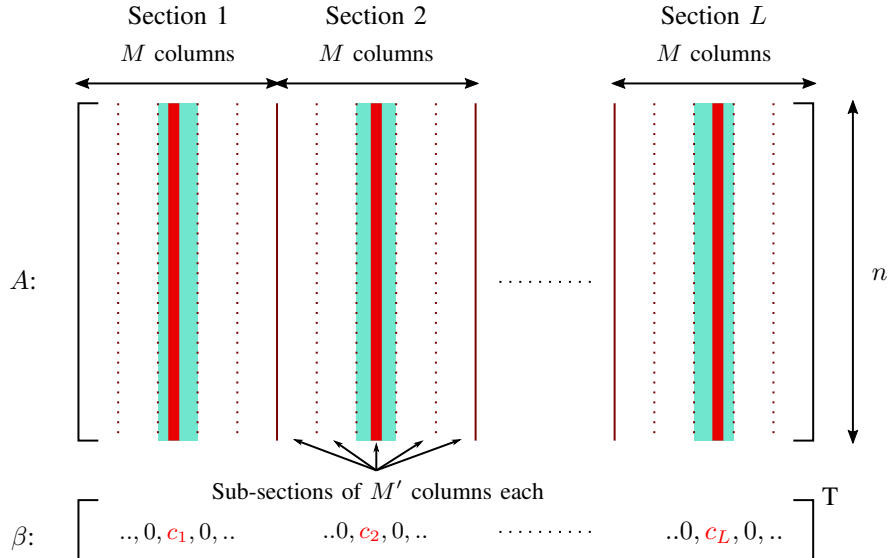
	\vspace{-0.15in}
	\caption{Random binning with SPARCs. Each section is divided into sub-sections of $M'$ columns. Choosing one sub-section from each section forms a bin, i.e. a smaller SPARC. }
	\label{fig:nestedSPARC}
\end{figure}

A particularly useful characteristic of SPARCs for multi-user information theory problems and rate-splitting applications like secret key agreement is their inherent ability to perform binning through a nested code structure \cite{venk_12_aller}.
Consider a SPARC $\mathcal{C}$ with parameters $L, M, n$ designed for a rate $R$ such that $L \log M = nR$. Each of the $L$ sections in the dictionary $ \mathbf{A}$ and correspondingly in the coefficient vector $\beta$, has $M$ possible column choices. We can partition these $M$ columns within each section into $M/M'$ sub-sections, where each sub-section contains $M'$ columns, where  $M' \mid M$.
A bin is formed by selecting one such sub-section from each of the $L$ sections. The codewords within this bin are those SPARCs formed by choosing one column from the selected sub-section in each of the $L$ sections. This results in a smaller SPARC codebook, effectively a sub-code, with $(M')^L$ codewords. The total number of such distinct bins is $(M/M')^L$.
Recognizing $R-R'$ as the rate of bin indices, where $n(R-R') = L \log(M/M')$, and $nR' = L \log M'$ as the rate of the sub-codebook within each bin, then the overall code of rate $R$ is partitioned into $e^{n(R-R')}$ bins, each containing $e^{nR'}$ codewords. This structure is illustrated in Figure \ref{fig:nestedSPARC}.
Formally, a nested SPARC is defined as follows
\begin{defn}
	A nested SPARC $\mathcal{C}(R,R',n,\alpha,\alpha')$ is defined by an $n \times ML$ dictionary $\mathbf{A}$ with i.i.d. $\mathcal{N}(0,1/n)$ entries. The parameters are such that $M=L^\alpha$ and $M'=L^{\alpha'}$ for section size rates $\alpha, \alpha' > 0$ and $\alpha L\log L=nR$.  The overall code has rate $R$, and it is divided into $e^{n(R-R')}$ bins. Each bin is a smaller SPARC codebook of rate $R'$. The number of columns $M'$ in each sub-section is determined by $M'^L = e^{nR'}$.
\end{defn}

\subsection{Achieving the WZ rate}
The nested structure of SPARCs makes them well-suited for implementing the WZ coding scheme \cite{WynerZiv76}, which addresses lossy source coding with side information available only at the decoder. This scenario is central to the information reconciliation step in our secret key agreement protocol.
Consider an i.i.d. Gaussian source $X \sim \mathcal{N}(0,\sigma_X^2)$ that needs to be compressed. The decoder has access to non-causal side information $Y$ correlated with $X$. In our SK context, $Y = X + \eta_b$, where $\eta_b \sim \mathcal{N}(0, \sigma_{\eta_b}^2)$ is independent Gaussian noise. The WZ coding scheme using SPARCs, as detailed in \cite{venk_12_aller}, often involves an auxiliary random variable $U = X + V$, where $V \sim \mathcal{N}(0, Q)$ is Gaussian noise independent of $X$, introduced by Alice for quantization. Alice observes $X$ and effectively quantizes it by quantizing $U$.

\begin{enumerate}
	\item \textit{ Quantization of $U$}: Alice quantizes $U^n$ to a codeword $\mathbf{u}^n$ from a SPARC codebook of rate $R_1$. This process means $X$ is represented as $\xi U$, where $\xi = \sigma_X^2 / (\sigma_X^2+Q)$. The effective distortion for $X$ is 
	\begin{IEEEeqnarray}{rcl}
		\label{eq_effective_distortion}
		D_X = \mathbb{E}[(X-\xi U)^2] = \frac{\sigma_X^2 Q }{(\sigma_X^2+Q)}.
	\end{IEEEeqnarray} 
	Reliable quantization of $U$ requires 
	\begin{IEEEeqnarray}{rcl}
		\label{eq_reliable_quantization_rate}
	R_1 > R^*(D_X) = \frac{1}{2}\log\left(\frac{\sigma_X^2+Q}{Q}\right),
	\end{IEEEeqnarray} 
	achievable by Theorem 1. 
	\item \textit{Binning}: Using the nested SPARC structure $\mathcal{C}(R_1, R_2, ...)$, Alice identifies the bin index corresponding to her chosen codeword $\mathbf{u}^n$. She transmits this bin index (at public communication rate $R_P = R_1 - R_2$) to Bob.
	\item \textit{Decoding}: Bob uses the received bin index and his side information $Y^n$ to recover $\mathbf{u}^n$. The relationship between $Y$ and $U$ is $Y = X + \eta_b = \xi U + (X - \xi U) + \eta_b = \xi U + V' + \eta_b$, where $V' = X - \xi U$ is the effective quantization noise. Recovering $U$ from $Y$, given the bin, is a channel coding problem over an effective channel where $\xi U$ is the signal and $V'+\eta_b$ is the noise. The signal-to-noise ratio (SNR) for this effective channel is:
	\begin{align}
		\label{eq_snr}
		\gamma_{WZ} = \frac{\sigma_X^4}{\sigma_X^2 Q +(\sigma_X^2+Q)\sigma_Z^2}.
	\end{align}
\end{enumerate}

	For Bob to successfully recover $\mathbf{u}^n$ with high probability, the binning rate $R_2$ must satisfy 
	\begin{IEEEeqnarray}{rcl}
	\label{eq_wz_capacity}
		R_2 < C_{WZ} = \frac{1}{2}\log(1 + \gamma_{WZ}),
	\end{IEEEeqnarray} 
	where $C_{WZ}$ is the capacity of the effective channel between $U$ and $Y$. This is guaranteed if the number of codewords in each bin satisfies $M'^L < e^{n C_{WZ}}$. The overall performance is summarized in the following theorem, adapted from \cite[Thm. 2]{venk_12_aller}.

\begin{thm}\cite{venk_12_aller} \label{Wynerziv_SPARCs}
	For an i.i.d. Gaussian source $X \sim \mathcal{N}(0, \sigma_X^2)$, side information $Y = X + \eta_b$ with $\eta_b \sim \mathcal{N}(0, \sigma_{\eta_b}^2)$, and a target distortion $D_X$ for reconstructing $X$ via $\xi U$:
	If the quantization rate $R_1 > R^*(D_X)$ and the inner bin rate $R_2 < C_{WZ}$, then there exists a sequence of nested SPARCs $\{\mathcal{C}_n\}$ such that $X^n$ can be quantized to $\xi \mathbf{u}^n$ with average distortion at most $D_X$. Furthermore, $\mathbf{u}^n$ can be reliably recovered by the decoder from the bin index and the side information $Y^n$ with vanishing error probability as $n \to \infty$. The existence of such SPARCs requires the section size rate parameter $\alpha$ for the overall code to satisfy
	\begin{align}
		\label{eq_alpha}
		\alpha > \max \biggl\{\frac{2.5R_1}{R_1-\sigma_X^2/(\sigma_X^2+Q)}, \frac{R_1}{R_2}\alpha_0(\gamma_{WZ})\biggr\}
	\end{align}
	where $\alpha_0(v)$ is given by
	\begin{equation}
		\alpha_0(v) =
		\begin{cases}
			\displaystyle \frac{4v(1+v)\log(1+v)}{\big[(1+v)\log(1+v) - v\big]^2}, & \text{if } v < v^*, \\[10pt]
			\displaystyle \frac{(1+v)\log(1+v)}{(1+v)\log(1+v) - 2v}, & \text{if } v \geq v^*,
		\end{cases}
		\label{eq:alpha0}
	\end{equation}
	and $v^*$ is the solution of $(1+v^*)\log(1+v^*) = 3v^*$.
\end{thm}

\section{Secret key agreement}\label{sec_ska}
This section first introduces the system model for secret key agreement from correlated Gaussian sources, defining the roles of the legitimate parties and the eavesdropper, along with the communication constraints. We then formally define the objectives of a key-distillation strategy and the associated performance metrics. Finally, we outline the general structure of our proposed SPARC-based secret key agreement protocol, detailing the three main operational steps.

\subsection{System Model}
We consider an i.i.d. memoryless source model for secret-key agreement. Three terminals are involved: Alice (the first legitimate party), Bob (the second legitimate party), and Eve (the eavesdropper). Alice observes a sequence $X^n = (X_1, ..., X_n)$, where $X_i \sim \mathcal{N}(0, \sigma_X^2)$, whereas Bob observes a sequence $Y^n = (Y_1, ..., Y_n)$, where $Y_i = X_i + \eta_{b,i}$. The noise sequence $\eta_b^n$ consists of i.i.d. components $\eta_{b,i} \sim \mathcal{N}(0, \sigma_{\eta_b}^2)$, independent of $X^n$. On the other side, Eve observes a sequence $Z^n = (Z_{1}, ..., Z_{n})$, where $Z_i = X_i + \eta_{e,i}$. The noise sequence $\eta_e^n$ consists of i.i.d. components $\eta_{e,i} \sim \mathcal{N}(0, \sigma_{\eta_e}^2)$, independent of $X^n$ and $\eta_b^n$. The components ($X_i, Y_i, Z_i$) are thus jointly Gaussian with zero mean. Alice and Bob aim to agree on a common secret key K by processing their respective observations $X^n$ and $Y^n$. They can communicate over a public, noiseless, and authenticated channel. All messages M exchanged over this public channel are also fully observed by Eve. Eve's goal is to gain information about the secret key $K$.

\subsection{Key-Distillation Strategy and Performance Metrics}
A key-distillation strategy $\mathcal{S}_n$ for a blocklength $n$ is defined by the following components:
\begin{itemize}
	\item A key alphabet $\mathcal{K}_n = \{1, ..., |\mathcal{K}_n|\}$, where $|\mathcal{K}_n| = e^{n R_K}$ and $R_K$ is the secret key rate in nats per source symbol.
	\item An alphabet $\mathcal{M}_n$ for the public messages exchanged, where $|\mathcal{M}_n| \approx e^{n R_P}$ and $R_P$ is the public communication rate. For simplicity, we assume Alice sends a single message $M$ to Bob.
	\item An encoding function at Alice: $\phi_n: \mathcal{X}^n \to \mathcal{M}_n$, which generates the public message $M = \phi_n(X^n)$.
	\item A key generation function at Alice: $\kappa_A^{(n)}: \mathcal{X}^n \to \mathcal{K}_n$, which generates Alice's key $K_A = \kappa_A^{(n)}(X^n)$.
	\item A key generation function at Bob: $\kappa_B^{(n)}: \mathcal{Y}^n \times \mathcal{M}_n \to \mathcal{K}_n$, which generates Bob's key $K_B = \kappa_B^{(n)}(Y^n, M)$.
\end{itemize}
The performance of a sequence of key-distillation strategies $\{\mathcal{S}_n\}_{n=1}^\infty$ is evaluated based on three main criteria:
\begin{enumerate}
	\item \textit{Reliability}: Alice and Bob must agree on the same key with high probability. This is measured by the average error probability, that is
	\begin{align}
		\label{reliability_SKA}
		\mathbb{P}_e(\mathcal{S}) = \mathbb{P}\left(K_A \neq K_B\right),
	\end{align} 
	A strategy is reliable if $\lim_{n\to\infty} P_e(\mathcal{S}_n) = 0$. 
	\item \textit{Secrecy}: The generated key $K$, assuming $K_A=K_B=K$, must be nearly independent of the eavesdropper's total information, which includes her observation $Z^n$ and the public message $M$. A strong secrecy criterion is often defined using the variational distance between the joint distribution $P_{K,M,Z^n}$ and the product of marginals $P_K P_{M,Z^n}$, that is
		\begin{IEEEeqnarray}{rcl}
		\nonumber
		&\mu_n(\mathcal{S}_n) = \frac{1}{2} \| P_{K,M,Z^n} - P_K P_{M,Z^n} \|_1 \\
		\label{secrecy_crit}
		&= \frac{1}{2} \sum_{k,m,z^n} |P_{K,M,Z^n}(k,m,z^n) - P_K(k)P_{M,Z^n}(m,z^n)|. \Dsupersqueezeequ
	\end{IEEEeqnarray}
	Strong secrecy is achieved if $\lim_{n\to\infty} \mu_n(\mathcal{S}_n) = 0$. This implies that the information leakage to Eve, $I(K; M, Z^n)$, also vanishes when normalized by $n$.
	\item \textit{Key Uniformity}: The generated key $K$ should be close to uniformly distributed over the key alphabet $\mathcal{K}_n$. This is measured by:
	\begin{align}
		\label{uniformity_SKA}
		\mathbb{U}(\mathcal{S}) = \log|\mathcal{K}| - H(K).
	\end{align}
	Perfect uniformity implies $U(\mathcal{S}_n) = 0$. Strong secrecy typically implies that $(1/n)H(K) \to R_K$, ensuring asymptotic uniformity. Additionally, the metric $T(S) = \log|\mathcal{K}| - H(K|M,Z^n)$, which is called security index~\cite{bloch_book}, is maximized when strong secrecy holds. 	
\end{enumerate}

A secret key rate $R_K$ is said to be achievable with a public communication rate $R_P$ if there exists a sequence of strategies $\{\mathcal{S}_n\}$ such that $\liminf_{n\to\infty} (1/n)\log|\mathcal{K}_n| \ge R_K, \limsup_{n\to\infty} (1/n)\log|\mathcal{M}_n| \le R_P$, and the conditions for reliability, strong secrecy, and key uniformity are all satisfied.

\subsection{Proposed SPARC-Based Protocol Structure}
Our secret key agreement protocol, building upon the general framework established in prior works for Gaussian sources~ \cite{Nitinawarat_12, Watanabe_10}, consists of three main operational steps, specifically implemented using the SPARC framework:

\begin{enumerate}
	\item \textit{Source Quantization}:
	Alice observes $X^n$ and aims to quantize it to a representation that can be reliably communicated to Bob. To facilitate WZ coding, she considers an auxiliary Gaussian random variable $U = X + V$, where $V \sim \mathcal{N}(0, Q)$ is quantization noise independent of $X$, with variance $Q$ being a design parameter. Alice uses a rate-$R_1$ SPARC codebook to find a codeword $\mathbf{u}^n$ such that $X^n$ is effectively quantized to $\xi \mathbf{u}^n$ with a mean-squared distortion given in~\eqref{eq_effective_distortion}, i.e. $R_1>R^*(D_X)$.
	\item \textit{Information Reconciliation}:
	Having selected $\mathbf{u}^n$, Alice uses the nested structure of her rate-$R_1$ SPARC codebook, partitioned into bins of sub-rate $R_2 < R_1$, to determine the bin index $M$ corresponding to $\mathbf{u}^n$. Alice transmits this message $M$ over the public channel to Bob. The rate of this public communication is $R_P = R_1 - R_2$.
	Bob receives $M$ and using his side information $Y^n$, attempts to decode $\mathbf{u}^n$ from the specified bin. This is a WZ decoding problem, and successful recovery is possible if $R_2 < C_{WZ}$. Upon successful decoding, Bob shares the common sequence $\mathbf{u}^n$ with Alice.
	\item \textit{Privacy Amplification}:
	Alice and Bob both possess the common sequence $\mathbf{u}^n$. However, Eve has her own correlated observation $Z^n$ and has also observed the public message $M$. Therefore, $\mathbf{u}^n$ is generally not yet secure enough to be used directly as a key. To extract a secure key, Alice and Bob apply the same randomly chosen universal hash function $\kappa$ (from a publicly known family of hash functions) to $\mathbf{u}^n$ to generate the final secret key: $K = K_A = K_B = \kappa(\mathbf{u}^n)$. The length of the key $K$, and thus the rate $R_K$, is chosen such that $R_K$ is less than the conditional entropy of $U^n$ given Eve's information, effectively hashing out Eve's knowledge. This step ensures both secrecy and uniformity of the final key $K$.
\end{enumerate}

\section{Main Results}\label{sec_main}
\label{sec:main_results}
This section presents the core theoretical contributions of this work, establishing the achievable performance of the proposed SPARC-based secret key agreement protocol in the asymptotic limit of the blocklength. We detail the SPARC parameterization, state the key theorems regarding achievable rates and strong secrecy guarantees, and analyze the protocol's optimality gap relative to fundamental information-theoretic limits.

\subsection{Protocol Implementation Parameters with SPARCs}
\label{subsec:protocol_params}
The three-step secret key agreement protocol detailed in Section III is implemented using a nested SPARC structure. The rates for different stages are assigned to leverage the optimal properties of SPARCs as follows:

\begin{enumerate}
	\item \textit{Source Quantization:} Alice employs an overall SPARC codebook $\mathcal{C}_1$ of rate $R_1$ to quantize her source $X^n$ to $\xi \mathbf{u}^n$ via the auxiliary variable $U=X+V$. The parameter $Q = \mathrm{Var}(V)$ controls the distortion $D_X$. From Theorem 1, reliable quantization requires the SPARC rate $R_1$ (nats per symbol) to satisfy:
	\begin{align} 
		\label{eq:R1_condition_main_final}
		R_1 > R^*(D_X) = \frac{1}{2}\ln\left(\frac{\sigma_X^2+Q}{Q}\right) 		
	\end{align}	
	\item \textit{Information Reconciliation:} The codebook $\mathcal{C}_1$ is structured as a nested code, effectively $\mathcal{C}(R_1, R_2)$. Alice transmits the bin index $M$ corresponding to $\mathbf{u}^n$, which identifies a sub-codebook (bin) $\mathcal{C}_{bin}(M)$ of rate $R_2$. The public communication rate is $R_P = R_1 - R_2$. For Bob to reliably decode $\mathbf{u}^n$ from this bin using his side information $Y^n$, the inner bin rate $R_2$  (nats per symbol) must be less than the WZ capacity $C_{WZ,Bob}$ of the effective channel between $U$ and $Y$ as established by Theorem 2:
	\begin{equation}
		\label{eq:R2_condition_main_final}
		R_2 < C_{WZ,Bob} = \frac{1}{2}\ln(1 + \gamma_{WZ,Bob}),
	\end{equation}
	where 
	\begin{align}
		\label{eq_bob_wz_snr}
			\gamma_{WZ,Bob} = \frac{\sigma_X^4}{\sigma_X^2 Q + (\sigma_X^2+Q)\sigma_{\eta_b}^2}.
	\end{align}

	 Consequently, the minimum public communication rate is:
	\begin{align}
		\label{eq:Rp_condition_main_final}
		R_P > R^*(D_X) - C_{WZ,Bob}
	\end{align}
	The construction of such SPARCs, including the choice of section size rate $\alpha$, is guaranteed by Theorem 2.
	\item \textit{Privacy Amplification:} Alice and Bob apply a 2-universal hash function $\kappa_n$ to their agreed-upon common sequence $u^n$ to generate the secret key $K = \kappa_n(u^n)$. The secret key rate $R_K$ is chosen to eliminate Eve's information about $u^n$, thereby ensuring strong secrecy.
\end{enumerate}

\subsection{Achievable Rates and Strong Secrecy Guarantees}
\label{subsec:achievable_rates_secrecy_text}

The security of any secret key agreement protocol hinges on two main aspects: ensuring the legitimate parties agree on the same information, and ensuring this information is incomprehensible to an eavesdropper. Our protocol uses SPARCs for the agreement phase and universal hashing for the security phase. To formally establish the strong secrecy of the final key $K$ derived from $U^n$, we consider the following. Let $U^n$ be the sequence shared by Alice and Bob after reconciliation, from which a key $K$ is generated. Let $(M, Z^n)$ be the total side information available to the eavesdropper, where $M=\phi_n(U^n)$ is the public message.

The uncertainty Eve has about $U^n$ is characterized by the conditional R\'enyi entropy of order 2 ~\cite{Renyi_entropy}. This is defined via the conditional collision probability for a given realization $(m,z^n)$:
	\begin{IEEEeqnarray}{rcl}
			\label{eq_collision_prob}
			P_{\mathrm{coll}}(U^n|m,z^n) = \sum_{u^n:\phi_n(u^n)} P(U^n=u^n|M=m,Z^n=z^n)^2. \Tsupersqueezeequ
		\end{IEEEeqnarray}
The asymptotic average conditional R\'enyi entropy rate is then
	\begin{IEEEeqnarray}{rcl}
			\label{eq_renyi_entopy_order_2}
		H_2(U|M,Z) = \lim_{n\to\infty} \frac{1}{n}\mathbb{E}[-\log_2 P_{\mathrm{coll}}(U^n|M,Z^n)]. \Tsupersqueezeequ
		\end{IEEEeqnarray} 
The strong secrecy proof relies on the assumption that the instantaneous conditional R\'enyi entropy concentrates around its mean. We say the entropy concentrates sufficiently fast if for any $\delta > 0$, the probability
	\begin{IEEEeqnarray}{rcl}
		\label{eq_pld_def}
		\mathbb{P}\left(\left| \frac{1}{n}\left(-\log_2 P_{\mathrm{coll}}(U^n|M,Z^n)\right) - H_2(U|M,Z) \right| > \delta \right)\Dsupersqueezeequ
	\end{IEEEeqnarray} 
decays exponentially with a large deviation exponent $E_{LD} > 0$. Setting~\eqref{eq_pld_def} equal to $P_{LD}(\delta)$, it shall holds that $P_{LD}(\delta) \le c_1 e^{-nE_{LD}}$ for some constant $c_1$.

\begin{thm}[Strong Secrecy via Universal Hashing]
	\label{thm:hashing_secrecy_main_results_text}
	 Let $\kappa_n: \mathcal{U}^n \to \mathcal{K}_n$ be a hash function chosen uniformly at random from a 2-universal family $\mathcal{H}_n$, producing a key $K = \kappa_n(U^n)$ of rate $R_K = (1/n)\ln|\mathcal{K}_n|$. Strong secrecy  is achieved if 
	 	\begin{equation}
	 		\label{eq_rk_condition_thm_1}
	 		R_K < H_2(U|M,Z) \ln 2 - \nu,
	 	\end{equation}
	 	for some $\nu > 0$ and if 
	 	\begin{equation}
		 		\label{eq_rk_condition_thm_2}
	 		E_{LD} > R_K
	 	\end{equation}
\end{thm}
\begin{IEEEproof}
	The proof is provided in Appendix A.
\end{IEEEproof}
This theorem provides the basis for privacy amplification. The R\'enyi entropy measures the uncertainty Eve has about $U^n$ even after observing the public message $M$ and her own correlated data $Z^n$. If the rate of the key $R_K$ is less than this uncertainty, the hashing process effectively rules out Eve's partial information, making the final key $K$ appear random and uncorrelated with her knowledge. The variational distance in~\eqref{secrecy_crit} approaching zero signifies this strong form of secrecy.

Building upon this general security principle, the next theorem is the central result for our SPARC-based secret key agreement protocol. It specifies the maximum rate at which Alice and Bob can generate such a strongly secret key, and the corresponding public communication cost, by leveraging the unique capabilities of SPARCs for the initial agreement on $U^n$. The key idea is that Alice and Bob must first agree on $U^n$ reliably, addressed by SPARCs' WZ optimality, and then the rate of the final secret key $K$ is limited by how much better Bob's information about $U^n$, via $Y^n$ and $M$, is compared to Eve's information via $Z^n$ and $M$.
\begin{thm}[Achievable Secret Key Rate]
	\label{thm:sk_rate_strong_secrecy_main_results_text}
	For the Gaussian source model and for any chosen auxiliary quantization noise variance $Q > 0$, the SPARC-based secret key agreement protocol can achieve any secret key rate $R_K$ and public communication rate $R_P$ pair satisfying:
	\begin{align}
		R_K < I(U; Y|M) - I(U; Z|M)
		\label{eq:new_thm4_achievable_Rk} 
	\end{align}
	and 
	\begin{align}
		R_P > I(U; X) - I(U; Y|M)
		\label{eq:new_thm4_achievable_Rp}  
	\end{align}
	where all mutual informations are evaluated for the joint distribution $P_{U,X,Y,Z,M}$ induced by the protocol. As $n \to \infty$, the generated key $K$ satisfies the criteria \eqref{reliability_SKA}, \eqref{secrecy_crit} and \eqref{uniformity_SKA}.
\end{thm}
\begin{IEEEproof}
	The proof is provided in Appendix B.
\end{IEEEproof}
This theorem quantifies the performance of the entire SPARC-based system.
Condition \eqref{eq:new_thm4_achievable_Rp} ensures that Alice sends enough public information $M$, at rate $R_P$, for Bob to reliably reconstruct $U^n$, given his side information $Y^n$. This is effectively the rate required for WZ source coding of $U^n$ with side information $Y^n$ at the decoder, where $M$ is the helper message.
Condition \eqref{eq:new_thm4_achievable_Rk} sets the upper limit on the secret key rate. The term $I(U; Y|M)$ represents the information Bob has about $U^n$, while $I(U; Z|M)$ is the information Eve has about $U^n$. The difference is essentially the information advantage Bob has over Eve regarding $U^n$. This advantage is converted into a secret key. If Bob can decode $U^n$ perfectly given $M$ and $Y^n$, then $H(U|M,Y^n) \approx 0$, and $I(U;Y|M) \approx H(U|M)$. The key rate then becomes $R_K < H(U|M) - I(U;Z|M) \approx H(U|M,Z)$. This directly connects to the premise of Theorem 3 for successful privacy amplification. The SPARC framework's ability to implement the reconciliation stage optimally via Theorem 1 and 2 allows these information-theoretic rates to be achieved.

\subsection{Operational Gaussian Rate Bounds}
The mutual information terms in Theorem 4 can be made more explicit for the Gaussian setting. Since $M=\phi_n(U^n)$ is a deterministic function of $U^n$, and Bob reliably decodes $U^n$ given $M$ and $Y^n$, implying $(1/n)H(U^n|M,Y^n) \to 0$, the condition on $R_K$ in \eqref{eq:new_thm4_achievable_Rk} simplifies asymptotically to $R_K < (1/n)H(U^n|M,Z^n)$.
Using the operational capacities $C_{\mathrm{WZ,Bob}}$ from \eqref{eq:R2_condition_main_final} and setting
\begin{align}
	\label{eq_eve_capacity}
	C_{\mathrm{WZ,Eve}} = \frac{1}{2}\ln(1 + \gamma_{\mathrm{WZ,Eve}}),
\end{align}
where
\begin{align}
	\label{eq_eve_snr}
	\gamma_{\mathrm{WZ,Eve}} = \frac{\sigma_X^4}{\sigma_X^2 Q + (\sigma_X^2+Q)\sigma_{\eta_e}^2}.
\end{align}
Then the achievable secret key rate is:
\begin{align}
	R_K < C_{\mathrm{WZ,Bob}} - C_{\mathrm{WZ,Eve}} = \frac{1}{2}\ln\left( \frac{1 + \gamma_{\mathrm{WZ,Bob}}}{1 + \gamma_{\mathrm{WZ,Eve}}} \right).
	\label{eq:Rk_gaussian_explicit_final_main_results_v3}
\end{align}
The public communication rate, using rates from \eqref{eq:R1_condition_main_final} and~\eqref{eq:R2_condition_main_final}, is:
\begin{align}
	R_P > R^*(D_X) - C_{\mathrm{WZ,Bob}} = \frac{1}{2}\ln\left( \frac{(\sigma_X^2+Q)/Q}{1 + \gamma_{\mathrm{WZ,Bob}}} \right).
	\label{eq:Rp_gaussian_explicit_final_main_results_v3}
\end{align}
A key insight from \eqref{eq:Rk_gaussian_explicit_final_main_results_v3} is that the secret key rate is determined by the difference in Bob's and Eve's abilities to resolve the common randomness $U^n$.

\subsection{Optimality Gap}

The fundamental information-theoretic upper bound on the secret key rate for the considered Gaussian source model is $R_{K,\mathrm{opt}} = I(X; Y) - I(X; Z)$ \cite{Nitinawarat_12, Watanabe_10}. Our SPARC-based protocol, due to the introduction of the auxiliary variable $U = X + V$ for quantization, necessarily incurs a rate gap relative to this optimum.

\begin{cor}[Rate Gap] 
	The secret key rate $R_K$ achievable by the proposed SPARC protocol is separated from the optimal rate $R_{K,\mathrm{opt}}$ by a gap $\Delta R_K = R_{K,\mathrm{opt}} - R_K$. This gap can be expressed as:
		\begin{IEEEeqnarray}{rcl}
			\Delta R_K \ge [I(X; Y) - C_{\mathrm{WZ,Bob}}] - [I(X; Z) - C_{\mathrm{WZ,Eve}}].
		\label{eq:rate_gap_main_final_v3} \Tsupersqueezeequ
	\end{IEEEeqnarray} 
\end{cor}
Since $C_{WZ,Bob} \approx I(U;Y|M)$ and $C_{WZ,Eve} \approx I(U;Z|M)$ when considering the information flow about $U$ given $M$, the gap $\Delta R_K \approx I(X; Y | U) - I(X; Z | U)$. This term represents the net loss in the secrecy advantage $I(X;Y) - I(X;Z)$ due to Alice's processing for creating $U$ from $X$, instead of using $X$ directly.
The rate gap $\Delta R_K$ identified in Corollary~1 is primarily a function of the auxiliary quantization noise variance $Q = \mathrm{Var}(V)$. This parameter acts as a fundamental tuning knob for the protocol, controlling the trade-off between the achievable secret key rate and the required public communication cost.

At one extreme, as $Q \to 0$ (fine quantization), the auxiliary variable $U$ becomes a near-perfect copy of the original source $X$. Consequently, the capacities of the effective WZ channels for Bob and Eve, $C_{\mathrm{WZ,Bob}}$ and $C_{\mathrm{WZ,Eve}}$, converge to the mutual information $I(X;Y)$ and $I(X;Z)$, respectively. This causes the rate gap $\Delta R_K$ to vanish, and the protocol's achievable key rate $R_K$ approaches the theoretical optimum, $R_{K,\mathrm{opt}}$. However, this optimality comes at a steep price: the rate-distortion function $R^*(D_X) = (1/2)\ln((\sigma_X^2+Q)/Q)$ approaches infinity, leading to an unbounded public communication rate $R_P$.

At the other extreme, as $Q \to \infty$ (coarse quantization), $U$ becomes largely independent of $X$, carrying negligible information about the source. As a result, both $C_{\mathrm{WZ,Bob}}$ and $C_{\mathrm{WZ,Eve}}$ tend to zero, causing the achievable secret key rate $R_K$ to vanish. In this regime, the public communication rate $R_P$ also tends to zero, as there is little information to convey.

This reveals a fundamental design trade-off, which is numerically illustrated in the following section. As we will see in Figure \ref{fig:q_contour_tradeoff}, the boundary of the achievable $(R_P, R_K)$ region is traced by varying $Q$, clearly showing that achieving a higher $R_K$ necessitates a higher $R_P$. It is important to emphasize that this trade-off is inherent to the chosen SK agreement strategy and that the SPARCs themselves implement the component tasks optimally for the given $U$. That is, Theorem~1 ensures $R^*(D_X)$ is achieved for quantization, and Theorem~2 ensures the WZ rate $C_{\mathrm{WZ,Bob}}$ is achieved for reconciliation. The observed rate gap is therefore a consequence of the protocol's structure, not a sub-optimality of the SPARC coding scheme itself. An interesting aspect of this result is that the entire protocol leverages the same underlying SPARC machinery for distinct information-theoretic tasks---source coding, WZ coding, and implicitly the channel coding for reconciliation---showcasing the versatility and power of the SPARC framework.

\section{Numerical Illustration of Rate Trade-offs and Practical Constraints}
\label{subsec:numerical_illustration_final_latex_individual}

To visually illustrate the theoretical trade-offs derived from our analysis and to explore the practical implications of the SPARC code construction, we present a series of numerical examples. These plots are based on the operational rate bounds in~\eqref{eq:Rk_gaussian_explicit_final_main_results_v3} and~\eqref{eq:Rp_gaussian_explicit_final_main_results_v3}. A key aspect of this analysis is the role of the SPARC section size rate parameter $\alpha$. While the asymptotic rate formulas depend on the quantization noise variance $Q$, the very existence of a SPARC to achieve these rates depends on satisfying condition \eqref{eq_alpha} from Theorem~2. This condition, which we denote as $\alpha > \alpha_{\mathrm{req}}(Q)$, links the code structure to the protocol's operational rates. As the quantization becomes finer ($Q \to 0$), the required quantization rate $R_1(Q)$ tends to infinity, which in turn causes $\alpha_{\mathrm{req}}(Q)$ to also approach infinity.

Therefore, for any practical system with a fixed, finite section size rate $\alpha$, there exists a minimum feasible quantization variance, $Q_{\mathrm{min}}$, below which the required rates are unachievable by the chosen SPARC family. This induces a feasible operating region for $Q$ and, as shown below, leads to a non-trivial optimization for maximizing the secret key rate. In the following plots, we assume a practical choice of $\alpha=6$ and a normalized source variance $\sigma_X^2 = 1$.

\begin{figure}[htb]
	\centering
	\includegraphics[width=0.7\columnwidth]{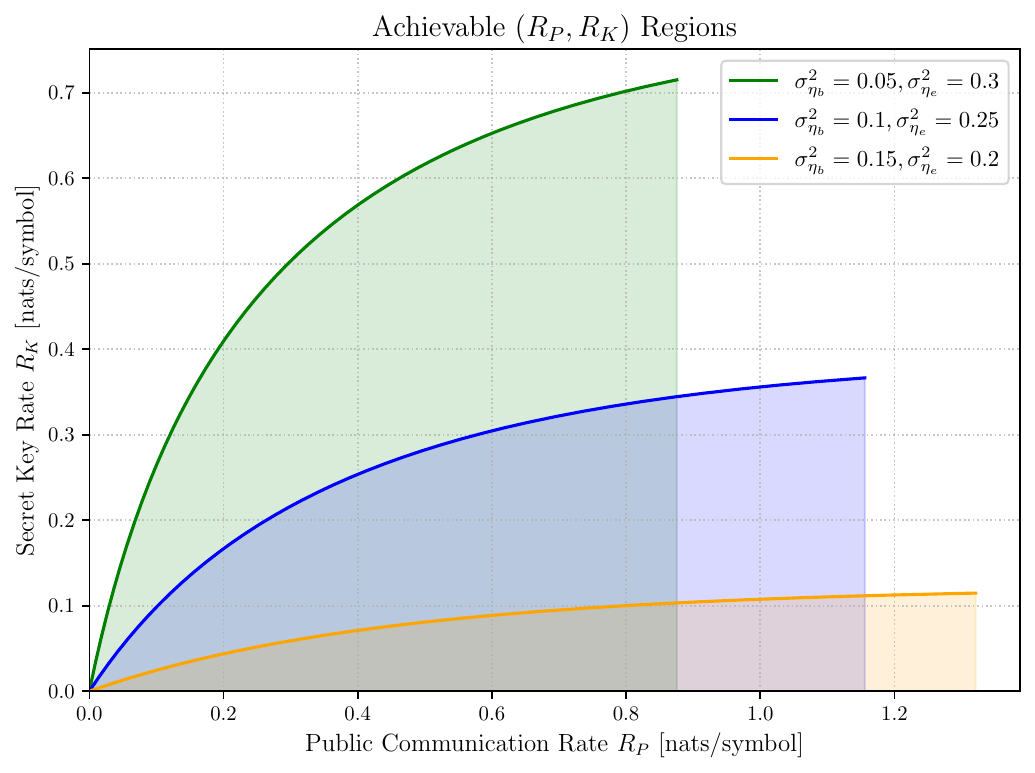}
	\caption{Achievable $(R_P, R_K)$ regions for three different scenarios of Bob's and Eve's channel qualities, generated by varying the quantization parameter $Q$. The size of the region is dictated by Bob's advantage over Eve.}
	\label{fig:rp_rk_regions}
\end{figure}

\begin{figure}[htb]
	\centering
	\includegraphics[width=0.7\columnwidth]{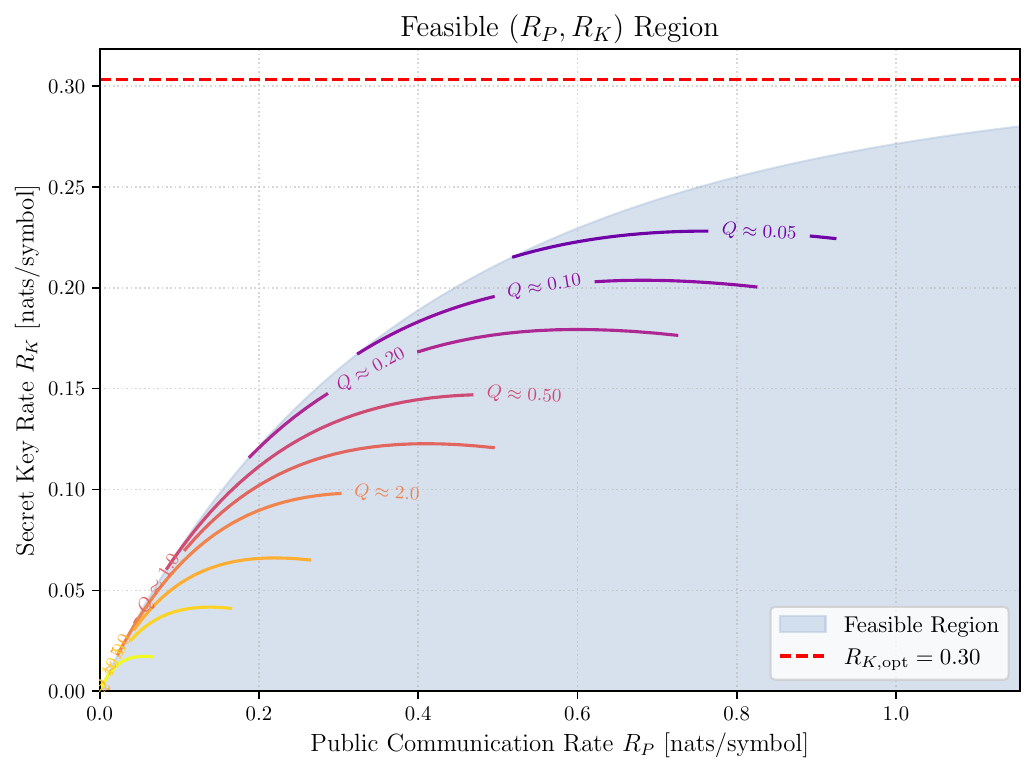}
	\caption{A detailed view of the achievable region for $\sigma_{\eta_b}^2=0.1$ and $\sigma_{\eta_e}^2=0.2$, with contour lines indicating the value of $Q$ that achieves each point on the boundary. The dashed line indicates the optimal secret key rate $R_{K,\mathrm{opt}}$.}
	\label{fig:q_contour_tradeoff}
\end{figure}

Figure \ref{fig:rp_rk_regions} depicts the entire achievable $(R_P, R_K)$ region for three scenarios with varying channel qualities for Bob and Eve. Each shaded region represents the set of rate pairs achievable by our protocol. The boundary of each region is traced by varying the quantization parameter $Q$ over its feasible range. This plot highlights a fundamental principle: the overall size of the achievable region is dictated by the secrecy advantage that Bob has over Eve. A larger gap between Bob's channel quality (lower $\sigma_{\eta_b}^2$) and Eve's (higher $\sigma_{\eta_e}^2$) results in a substantially larger achievable region.

A more detailed view of this trade-off is provided in Figure \ref{fig:q_contour_tradeoff} for a fixed scenario ($\sigma_{\eta_b}^2=0.1, \sigma_{\eta_e}^2=0.2$). The achievable boundary is shown with overlaid contour lines for the value of $Q$ required to operate at each point. This figure can be interpreted as a practical design guide. For instance, to achieve a high secret key rate approaching the optimum $R_{K,\mathrm{opt}}$, one must select a small value of $Q$ (e.g., $Q \approx 0.05$), which in turn demands a high public communication rate of $R_P \approx 0.8$ nats/symbol. Conversely, if public communication is constrained, for example to $R_P \approx 0.2$, the designer must accept a lower secret key rate of $R_K \approx 0.05$, which is achieved by choosing a coarser quantization with $Q \approx 5.0$. This numerically confirms the behavior predicted by our analysis of the rate gap in Corollary~1.

\begin{figure}[htb]
	\centering
	\includegraphics[width=0.7\columnwidth]{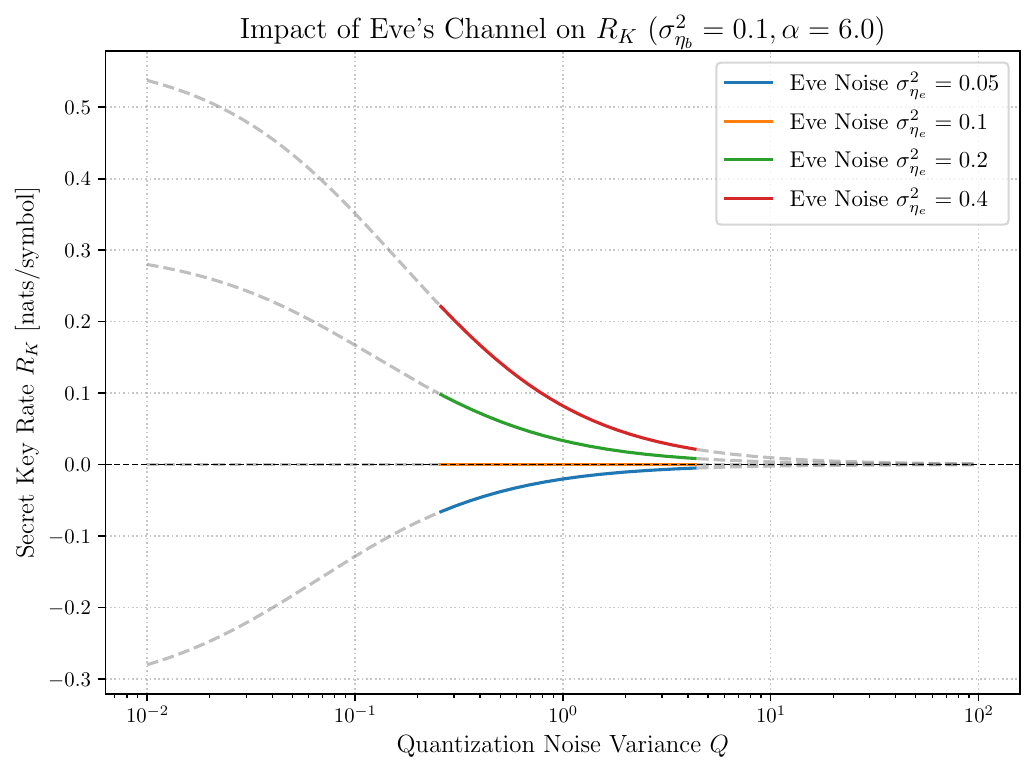}
	\caption{The impact of Eve's channel quality on $R_K$ for a fixed Bob's channel ($\sigma_{\eta_b}^2=0.1$) and fixed SPARC parameter $\alpha=6$. The non-monotonic behavior of $R_K$ as a function of $Q$ is due to the SPARC constraints.}
	\label{fig:eve_impact}
\end{figure}

\begin{figure}[htb]
	\centering
	\includegraphics[width=0.7\columnwidth]{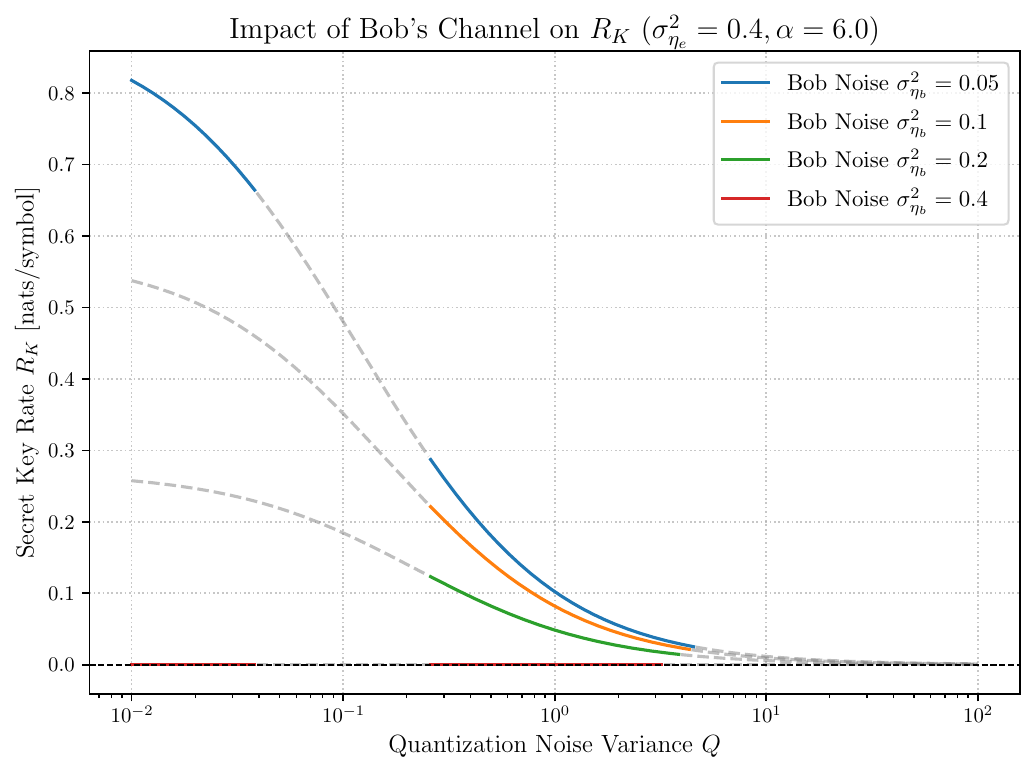}
	\caption{The impact of Bob's channel quality on $R_K$ for a fixed Eve's channel ($\sigma_{\eta_e}^2=0.4$) and fixed SPARC parameter $\alpha=6$.}
	\label{fig:bob_impact}
\end{figure}

The impacts of Eve's and Bob's channel qualities on the secret key rate $R_K$ are isolated in Figure \ref{fig:eve_impact} and Figure \ref{fig:bob_impact}, respectively. These plots reveal a key insight stemming from the SPARC constraints regarding the code structure. In Figure \ref{fig:eve_impact}, we observe that for any given Eve, there exists an optimal choice of $Q$ that maximizes $R_K$. This peak occurs because of the trade-off between two effects. As $Q$ decreases from a large value, the quality of the common randomness $U$ improves, increasing the potential secret key rate. However, as $Q$ becomes too small, the required SPARC section size rate, $\alpha_{\mathrm{req}}(Q)$, grows rapidly. For a fixed $\alpha=6$, values of $Q$ below a certain threshold become infeasible, causing the achievable rate to drop to zero, as indicated by the dashed gray lines. The peak of the solid curve thus represents the maximum achievable $R_K$ for the given SPARC family. As expected, a stronger eavesdropper (smaller $\sigma_{\eta_e}^2$) severely diminishes this peak rate.

Similarly, Figure \ref{fig:bob_impact} shows the same non-monotonic behavior for varying Bob's channel quality. A better channel for Bob (smaller $\sigma_{\eta_b}^2$) raises the entire achievable curve, enabling a higher maximum secret key rate. These plots demonstrate that the optimization of $Q$ in a practical SPARC-based system is a non-trivial constrained optimization problem. The problem can be formally stated as:
\begin{equation}
	Q_{\mathrm{opt}} = \arg\max_{Q > 0} \quad R_K(Q)
	\label{eq:q_optimization_problem}
\end{equation}
subject to :
\begin{align}
	\nonumber
	\alpha_{\mathrm{fixed}} &\ge \alpha_{\mathrm{req}}(Q) \\
	&:= \max\left\{ \frac{2.5R_1(Q)}{R_1(Q) - \frac{\sigma_X^2}{\sigma_X^2+Q}}, \frac{R_1(Q)}{R_2(Q)} \alpha_0(\gamma_{\mathrm{WZ,Bob}}(Q)) \right\},
	\label{eq:alpha_constraint_for_q}
\end{align}
where $R_1(Q)$, $R_2(Q)$, and $\gamma_{\mathrm{WZ,Bob}}(Q)$ are all functions of $Q$, and $\alpha_{\mathrm{fixed}}$ is the chosen section size rate for the SPARC family (e.g., $\alpha_{\mathrm{fixed}}=6$).

This optimization is non-trivial because the feasible set for $Q$, defined by the inequality in~\eqref{eq:alpha_constraint_for_q}, is not a simple interval and is determined by a complex, non-linear function. Since our analysis of the unconstrained objective function $R_K(Q)$ revealed a monotonic behavior (maximized as $Q \to 0$), the optimal solution to the constrained problem in~\eqref{eq:q_optimization_problem} must lie at the boundary of the feasible set. Specifically, the optimal $Q_{\mathrm{opt}}$ will be the minimum value of $Q$ that satisfies the constraint:
\begin{equation}
	Q_{\mathrm{opt}} = \inf \{ Q > 0 \mid \alpha_{\mathrm{fixed}} \ge \alpha_{\mathrm{req}}(Q) \}.
\end{equation}
This corresponds to finding the value of $Q$ that solves the equation $\alpha_{\mathrm{fixed}} = \alpha_{\mathrm{req}}(Q)$. Finding a closed-form analytical solution for this equation is challenging, but it can be readily solved numerically to find the optimal operating point for any given set of channel statistics and a chosen SPARC code structure defined by $\alpha_{\mathrm{fixed}}$. This highlights a crucial design consideration for practical implementations.

\section{Discussion and Conclusion}
\label{sec_coclusion}

\subsection{Complexity}
\label{subsec:complexity_discussion_final_as_is}
The SPARC encoding process itself is generally considered to be of low complexity. A SPARC codeword $\mathbf{x} = \mathbf{A}\beta$ is formed by selecting $L$ columns from the $n \times N$ dictionary matrix $\mathbf{A}$ (where $N=ML$) and performing a scaled sum. If the mapping from information bits to the $L$ active coefficients in $\beta$ is efficient, the dominant operation becomes $L$ vector additions of $n$-dimensional vectors, leading to a computational cost of approximately $O(nL)$. This compares favorably to schemes requiring dense matrix-vector multiplications if $L$ is significantly smaller than $N$.
The more critical aspect regarding complexity typically lies in SPARC decoding, or more generally, sparse signal recovery. In our SKA protocol, this arises in two main stages. Firstly, Alice's source quantization step, where $X^n$ is mapped to an optimal SPARC codeword $\mathbf{u}^n$ from the rate $R_1$ codebook $\mathcal{C}_1$, can be viewed as a nearest-neighbor search or a sparse representation problem. If approached as finding the best sparse coefficient vector $\beta_U$ that minimizes $||\mathbf{U}^n - \mathbf{A}\beta_U||^2$, the complexity is akin to that of SPARC decoding.
Secondly, and likely the dominant computational load, is Bob's WZ decoding step. Here, Bob must decode $\mathbf{u}^n$ from the restricted SPARC sub-codebook $\mathcal{C}_\mathrm{bin}(M)$ using his side information $Y^n$. This is a channel decoding task for a SPARC code over the effective WZ channel. Fortunately, several efficient algorithms exist for SPARC decoding. Approximate Message Passing (AMP) algorithms \cite{barron_1, RushGreigVenkataramanan17CapacityAchievingAMP, HsiehVenkataramanan21ModulatedSPARC} are iterative statistical decoders shown to achieve capacity for SPARCs with Gaussian dictionaries, typically with a complexity of $O(nN)$ per iteration, and often requiring a relatively small number of iterations. Greedy algorithms, such as Orthogonal Matching Pursuit (OMP) or the Match and Decode (MAD) algorithm tailored for SPARCs \cite{SinhaKannu23GSPARC}, offer alternative iterative approaches with complexities in the order of $O(LnN)$ or potentially faster with structured dictionaries. The binning inherent in Bob's WZ decoding is advantageous, as his search for each of the $L$ components of $\beta_U$ is reduced from $M$ columns to $M'$ columns, reducing the effective $N$ or the per-iteration cost for these algorithms. In contrast, convex optimization methods like Basis Pursuit, while robust, are generally more computationally intensive for large $N$.
The final privacy amplification step, which involves applying a 2-universal hash function to $\mathbf{u}^n$, is computationally very efficient, typically linear in the blocklength $n$, and does not constitute a bottleneck.

Overall, the computational complexity of the proposed SKA scheme is primarily determined by the SPARC decoding algorithm employed by Bob. The availability of efficient algorithms like AMP and MAD suggests that the scheme has the potential for practical implementation with manageable complexity, especially when considering that the parameters $L$ for sparsity and $M$ for the columns per section can be optimized. While a detailed comparative analysis with the complexity of other SKA schemes (e.g., those based on polar codes, which feature $O(n \log n)$ decoding, or lattice codes, which can involve computationally intensive lattice reduction or closest vector searches) is beyond the current scope, SPARC-based methods are recognized for their competitive performance and moderate complexity in many sparse recovery and communication applications. The choice of dictionary structure (e.g., random Gaussian vs. deterministic structured dictionaries) can also further influence the practical decoding speed.

\subsection{Conclusion}
In this paper, we have proposed and rigorously analyzed a secret key agreement protocol for correlated Gaussian sources that leverages the advantageous properties of SPARCs. Our approach systematically employs SPARCs for the critical stages of source quantization and information reconciliation. Privacy amplification is subsequently performed using 2-universal hash functions to ensure robust security.
The primary contribution of this work is the demonstration that this SPARC-based protocol can achieve secret key rates that approach the fundamental information-theoretic optimum $I(X;Y) - I(X;Z)$. We have explicitly characterized the rate gap to this optimum, showing that it is a consequence of the initial quantization step and can be made arbitrarily small by decreasing the quantization noise variance $Q$, albeit at the cost of an increased public communication rate $R_P$. A key finding is that the protocol achieves strong secrecy, as defined by a vanishing variational distance between the actual distribution of the key and public messages and the ideal one. This ensures that the generated key is asymptotically independent of the eavesdropper's observations, including the public messages.
Furthermore, we showed that the choice of the quantization parameter $Q$ is not a simple monotonic trade-off but a non-trivial constrained optimization problem. The practical constraints on the SPARC code parameters induce a peak in the achievable secret key rate, and we have characterized the solution to this problem. This highlights the interplay between asymptotic theory and practical code design constraints. Our analysis highlights the versatility of the SPARC framework, where the same underlying coding principles are effectively utilized for distinct information-theoretic tasks. This establishes SPARCs as a theoretically sound and viable alternative to other structured codes, such as polar codes or lattice codes, for constructing practical secret key generation schemes in Gaussian environments. The inherent low-complexity encoding and decoding algorithms associated with SPARCs further enhance their potential for practical implementations.

Several important avenues for future research emerge from this work. A paramount direction is the comprehensive analysis of the proposed protocol in the finite blocklength regime. This would involve characterizing the trade-offs between key rate, public communication rate, blocklength, and the achievable levels of reliability and secrecy. Key challenges in this domain include determining the finite blocklength error probability of SPARC-based WZ decoding and, crucially, deriving tight bounds on the conditional min-entropy considering the specific structure imposed by SPARC quantization and binning. Insights from recent advances in finite blocklength information theory \cite{PolyanskiyPoorVerdu10Bounds} and studies on short-blocklength SPARCs \cite{SinhaKannu23GSPARC} will be instrumental here.
Further research could also explore the optimization of SPARC parameters (e.g., dictionary design, $L, M, \alpha, Q$) specifically for maximizing the finite blocklength secret key rate. Additionally, extending this SPARC-based framework to multi-terminal secret key agreement scenarios, such as conference key agreement or key agreement over multiple-access or broadcast channels where the superposition and binning capabilities of SPARCs might offer distinct advantages, presents another fertile area for investigation. Finally, adapting the proposed source model protocol to channel-based models of secret key generation using SPARCs would broaden its applicability.


{\appendices
\section*{Appendix A: Proof of Theorem 3}
\label{app:proof_thm3_hashing_secrecy} 

The proof establishes strong secrecy by bounding the variational distance using properties of 2-universal hash functions and arguments related to the concentration of conditional R\'enyi entropy of order 2.

Let $P_{K,M,Z^n}$ be the actual joint distribution of the generated key $K$, public message $M$, and Eve's observation $Z^n$, where $K=\kappa_n(U^n)$. Let $Q_{K,M,Z^n} \equiv P_K^{\mathrm{unif}} P_{M,Z^n}$ represent the ideal distribution, where the key $K$ is uniformly distributed over its alphabet $\mathcal{K}_n$ and is independent of $(M,Z^n)$. The variational distance, which we aim to bound, is $\Delta(P_{K,M,Z^n}, Q_{K,M,Z^n}) = \frac{1}{2}\|P_{K,M,Z^n} - Q_{K,M,Z^n}\|_1$.

A standard result for hashing with a randomly chosen function $\kappa_n$ from a 2-universal family $\mathcal{H}_n$, \cite[Lemma A.6]{Renes_SKA}, provides a bound on the expected variational distance. This bound relates the security of the final key to the average collision probability of the source sequence $U^n$ conditioned on the eavesdropper's side information. Formally, the average variational distance satisfies:
\begin{align}
	\nonumber
	\bar{\mu}_n &= \mathbb{E}_{\kappa_n \in \mathcal{H}_n} \left[ \Delta(P_{K,M,Z^n}, Q_{K,M,Z^n}) \right] \\
	&\le \frac{1}{2} \sqrt{ |\mathcal{K}_n| \mathbb{E}_{M,Z^n} [P_{\mathrm{coll}}(U^n|M,Z^n)] - 1 },
	\label{eq:proof_appA_mu_n_sqrt_coll_minus_1}
\end{align}
where $|\mathcal{K}_n|$ is the size of the key alphabet and $P_{\mathrm{coll}}(U^n|M,Z^n)$ is the conditional collision probability defined in~\eqref{eq_collision_prob}. For this bound to be useful, the term under the square root must be non-negative. We can also use a slightly looser but simpler upper bound that neglects the -1 term:
\begin{equation}
	\bar{\mu}_n \le \frac{1}{2} \sqrt{ |\mathcal{K}_n| \mathbb{E}_{M,Z^n} [2^{-H_2(U^n|M,Z^n)}] },
	\label{eq:proof_appA_mu_n_h2_form_app}
\end{equation}
where we have expressed the collision probability using the conditional R\'enyi entropy of order 2 given in \eqref{eq_renyi_entopy_order_2}, for the specific realization of $(M,Z^n)$. The remainder of the proof focuses on bounding the expectation in \eqref{eq:proof_appA_mu_n_h2_form_app}.
%
%

Let $H_2(U|M,Z) = \lim_{n\to\infty} (1/n)\mathbb{E}[H_2(U^n|M,Z^n)]$ be the asymptotic average conditional R\'enyi entropy rate. For any $\delta > 0$, define the typical set $\mathcal{T}_{\delta}^{(n)}$ for $(M,Z^n)$ as those realizations where the instantaneous entropy rate is close to the average:
\begin{align}
	\nonumber 
	\mathcal{T}_{\delta}^{(n)} =   \Biggl\{ (m,z^n) : \Big|& \frac{1}{n}H_2(U^n|M=m,Z^n=z^n) \\
		\label{eq:proof_appA_typical_set_h2_app}
	&- H_2(U|M,Z) \Big| \le \delta \Biggr\}
\end{align}
This implies that for $(m,z^n) \in \mathcal{T}_{\delta}^{(n)}$, we have $H_2(U^n|m,z^n) \ge n(H_2(U|M,Z) - \delta)$.
The probability of the atypical set, $P_{LD}(\delta) = P((M,Z^n) \notin \mathcal{T}_{\delta}^{(n)})$, is assumed to decay sufficiently fast with $n$, i.e., $P_{LD}(\delta) \le c_1 e^{-nE_{LD}}$ for some positive constants $c_1$ and $E_{LD}$.


We split the expectation $\mathbb{E}_{M,Z^n} [2^{-H_2(U^n|M,Z^n)}]$ using the typical set in \eqref{eq:proof_appA_typical_set_h2_app}.
	\begin{IEEEeqnarray}{rcl}
		\nonumber	
	\mathbb{E}_{M,Z^n} [2^{-H_2(U^n|M,Z^n)}] &=& \sum_{(m,z) \in \mathcal{T}_{\delta}^{(n)}} P(m,z) 2^{-H_2(U^n|m,z)} \Tsupersqueezeequ \\ 
		\label{eq:proof_appA_exp_bound_h2_app_1}
	&+& \sum_{(m,z) \notin \mathcal{T}_{\delta}^{(n)}} P(m,z) 2^{-H_2(U^n|m,z)} \Tsupersqueezeequ \\ 
	\nonumber
	&\le& P((M,Z^n) \in \mathcal{T}_{\delta}^{(n)}) \cdot 2^{-n(H_2(U|M,Z) - \delta)} \Tsupersqueezeequ \\  
		\label{eq:proof_appA_exp_bound_h2_app_2}
	&+& P((M,Z^n) \notin \mathcal{T}_{\delta}^{(n)}) \cdot 1 \Tsupersqueezeequ  \\ 
	\label{eq:proof_appA_exp_bound_h2_app}
	&\le& 2^{-n(H_2(U|M,Z) - \delta)} + P_{LD}(\delta),	 \Tsupersqueezeequ
	\end{IEEEeqnarray} 
	where equality in \eqref{eq:proof_appA_exp_bound_h2_app_1} is obtained by partitioning the expectation over the entire sample space of $(M,Z^n)$ into two disjoint sets; inequality in \eqref{eq:proof_appA_exp_bound_h2_app_2} follows from applying an upper bound to each term in the sum. For realization within the typical set $\mathcal{T}_{\delta}^{(n)}$, we use the lower bound on the instantaneous entropy rate from its definition in ~\eqref{eq:proof_appA_typical_set_h2_app}, which gives $H_2(U^n|m,z^n) \ge n(H_2(U|M,Z) - \delta)$, and thus $2^{-H_2(U^n|m,z^n)} \le 2^{-n(H_2(U|M,Z) - \delta)}$. For realizations in the atypical set, we use the trivial bound that any collision probability is less than or equal to one, which implies $H_2(U^n|m,z^n) \ge 0$ and therefore $2^{-H_2(U^n|m,z^n)} \le 1$.
	Finally, the inequality in \eqref{eq:proof_appA_exp_bound_h2_app} is due to the fact that $P((M,Z^n) \in \mathcal{T}_{\delta}^{(n)}) \le 1$ and by substituting \eqref{eq_pld_def}.

Substituting~\eqref{eq:proof_appA_exp_bound_h2_app} into~\eqref{eq:proof_appA_mu_n_h2_form_app}, and recalling $|\mathcal{K}_n| = e^{nR_K}$:
\begin{align*}
	\bar{\mu}_n &\le \frac{1}{2} \sqrt{ e^{nR_K} \left( 2^{-n(H_2^{\mathrm{bits}}(U|M,Z) - \delta)} + P_{LD}(\delta) \right) } \\
	&= \frac{1}{2} \sqrt{ e^{nR_K} e^{-n(H_2^{\mathrm{nats}}(U|M,Z) - \delta \ln 2)} + e^{nR_K} P_{LD}(\delta) } \\
	&= \frac{1}{2} \sqrt{ e^{n(R_K - (H_2^{\mathrm{nats}}(U|M,Z) - \delta \ln 2))} + e^{nR_K} P_{LD}(\delta) }
\end{align*}
where the R\'enyi entropy $H_2(U|M,Z)$ is in bits, denoted by $H_2^{\mathrm{bits}}(U|M,Z)$ and  $H_2^{\mathrm{nats}}(U|M,Z) = H_2^{\mathrm{bits}}(U|M,Z) \ln 2$. 

For $\bar{\mu}_n \to 0$ as $n \to \infty$, both terms under the square root must vanish. The first term vanishes if $R_K - H_2^{\mathrm{nats}}(U|M,Z) + \delta \ln 2 < 0$. The theorem condition is $R_K < H_2^{\mathrm{nats}}(U|M,Z) - \nu$ for some $\nu > 0$. We can choose $\delta \ln 2 = \nu/2$. Then the exponent is $R_K - H_2^{\mathrm{nats}}(U|M,Z) + \nu/2$. Since $R_K - H_2^{\mathrm{nats}}(U|M,Z) < -\nu$, the exponent is $< -\nu + \nu/2 = -\nu/2 < 0$. Thus, the first term vanishes exponentially with $n$.
Next, given $P_{LD}(\delta) \le c_1 e^{-nE_{LD}}$. The second term is $c_1 e^{n(R_K - E_{LD})}$. This term vanishes if $R_K < E_{LD}$. The theorem assumes that $P_{LD}(\delta)$ decays sufficiently fast, meaning that the exponent $E_{LD}$ is greater than the chosen key rate $R_K$. Since both terms under the square root vanish under the stated conditions \eqref{eq_rk_condition_thm_1}, \eqref{eq_rk_condition_thm_2}, and the condition on $P_{LD}(\delta)$), then $\bar{\mu}_n \to 0$ as $n \to \infty$.

The convergence $\bar{\mu}_n \to 0$ for the average over the hash family $\mathcal{H}_n$ implies that for any sequence $\zeta_n \to 0$, there exists a deterministic sequence of hash functions $\{\kappa_n^*\}_{n=1}^\infty$ from $\mathcal{H}_n$ such that the actual variational distance $\mu_n(\kappa_n^*) \le \zeta_n$ for sufficiently large $n$. This follows from Markov's inequality, ensuring that most hash functions in the family achieve good performance.

This completes the proof of Theorem 3.

\section*{Appendix B: Proof of Theorem 4}
\label{app:proof_thm4_sk_rate} 

The proof combines the reliability of the SPARC-based reconciliation phase (established by Theorems 1 and 2) with the security guarantees of the privacy amplification phase (established by Theorem 3). We prove that the protocol satisfies the three required conditions of reliability, strong secrecy, and key uniformity, and we justify the public communication rate.

	(i)\textit{Reliability:} The condition is that the probability of key disagreement vanishes, i.e., $P(K_A \neq K_B) \to 0$ as $n \to \infty$. The protocol is reliable if Alice and Bob can agree on the common sequence $U^n$ with vanishing error probability. Let $U_A^n$ be Alice's quantized sequence and let $\hat{U}_B^n$ be the output of Bob's WZ decoder. The key disagreement event $K_A \neq K_B$ occurs if $\hat{U}_B^n \neq U_A^n$, as both parties apply the same deterministic hash function $\kappa_n^*$.
	
	The reconciliation process consists of two stages. First, Alice quantizes $X^n$ to $U_A^n$ using a rate-$R_1$ SPARC. By Theorem 1, if Alice chooses $R_1 > R^*(D_X)$, this quantization can be performed reliably. Second, Bob decodes $U_A^n$ from the bin index $M$ and his side information $Y^n$. By Theorem 2, if the inner bin rate $R_2 < C_{\mathrm{WZ,Bob}}$, the probability of a decoding error, $P_{err,WZ} = P(\hat{U}_B^n \neq U_A^n)$, vanishes as $n \to \infty$.
	
	Therefore, the probability of key disagreement is bounded by the decoding error probability:
	\begin{equation}
		P(K_A \neq K_B) \le P(\hat{U}_B^n \neq U_A^n) = P_{err,WZ}.
	\end{equation}
	Since $P_{err,WZ} \to 0$ as $n \to \infty$, the reliability condition is satisfied.
	
	(ii) \textit{Strong Secrecy:} The condition is that the average variational distance vanishes, i.e., $\bar{\mu}_n \to 0$ as $n \to \infty$. This follows by showing that the achievable secret key rate from Theorem 4 satisfies the premise of Theorem 3 for a sufficiently large blocklength $n$. The main condition of Theorem 3 is that the secret key rate $R_K$ must satisfy:
	\begin{equation}
		R_K < H_2(U|M,Z) \ln 2 - \nu, \quad \text{for some } \nu > 0.
		\label{eq:proofB_thm3_condition_app}
	\end{equation}
	The achievable key rate from Theorem 4 is given by $R_K < I(U;Y|M) - I(U;Z|M)$. By the definition of conditional mutual information, this rate can be rewritten as:
	\begin{IEEEeqnarray}{rcl}
		\nonumber
		I(U;Y|M) - I(U;Z|M) &=& [H(U|M) - H(U|M,Y)]  \IEEEeqnarraynumspace \squeezeequ\\
		\label{Eq_Gauss_sigma_bound_proof_16}
		&-& [H(U|M) - H(U|M,Z)] \IEEEeqnarraynumspace \squeezeequ\\
		\label{Eq_Gauss_sigma_bound_proof_17}
		&=& H(U|M,Z) - H(U|M,Y), \IEEEeqnarraynumspace \squeezeequ
	\end{IEEEeqnarray}
	where \eqref{Eq_Gauss_sigma_bound_proof_16}  follows directly from the definition of conditional mutual information and \eqref{Eq_Gauss_sigma_bound_proof_17}  is obtained by canceling the common terms. 
	From the reliability analysis in part (i), Bob's reliable decoding of $U^n$ implies, by Fano's inequality, that the conditional entropy $(1/n)H(U^n|M,Y^n) \to 0$. Thus, the achievable key rate condition from Theorem 4 becomes asymptotically equivalent to $R_K < H^{\mathrm{nats}}(U|M,Z)$.
	
	It is a known property that the R\'enyi entropy of order 2 is a lower bound on the Shannon entropy, and for the i.i.d. source model considered, the asymptotic rates are equal: $H_2(U|M,Z) \ln 2 \approx H^{\mathrm{nats}}(U|M,Z)$ for large $n$. Therefore, if we choose a rate $R_K$ such that $R_K < H^{\mathrm{nats}}(U|M,Z) - \nu'$ for some small $\nu' > 0$, this choice will also satisfy the condition in \eqref{eq:proofB_thm3_condition_app} for a suitable $\nu$ and sufficiently large $n$. The second condition of Theorem 3, concerning the concentration of R\'enyi entropy, is a standard assumption for i.i.d. models. Thus, the premises of Theorem 3 are met, guaranteeing that $\bar{\mu}_n \to 0$.
	
	(iii) \textit{Key Uniformity:} The condition is that $(1/n)H(K) \to R_K$ as $n \to \infty$. This is a direct consequence of the strong secrecy condition. A property of the variational distance is that if $\bar{\mu}_n \to 0$, then the distribution of the generated key, $P_K$, must converge to the uniform distribution, $P_K^{\mathrm{unif}}$. Specifically, since marginalization does not increase the variational distance, $(1/2)\|P_K - P_K^{\mathrm{unif}}\|_1 \to 0$. This convergence in distribution implies the convergence of entropies, ensuring that $(1/n)H(K) \to (1/n)\ln|\mathcal{K}_n| = R_K$.
	
	Finally, we justify the condition on the public communication rate, $R_P > I(U;X) - I(U;Y|M)$. This arises directly from the rates required for the SPARC-based WZ coding implementation. Alice must describe the sequence $U^n$ to Bob. The rate of her initial SPARC codebook, $R_1$, must be greater than the rate-distortion function $R^*(D_X)$, which is approximately $I(U;X)$ for an appropriate choice of distortion $D_X$. Bob receives a message $M$ and uses side information $Y^n$. The rate of the bin codebook from which Bob decodes, $R_2$, must satisfy $R_2 < C_{\mathrm{WZ,Bob}}$, where $C_{\mathrm{WZ,Bob}} \approx I(U;Y|M)$ is the capacity for Bob to decode $U$ given $Y$ and knowledge of the bin specified by $M$. The public rate is $R_P = R_1 - R_2$. Substituting the bounds gives $R_P > I(U;X) - I(U;Y|M)$, which is consistent with the rate required for this helper-dependent source coding problem.

This completes the proof of Theorem 4.
}

\bibliographystyle{IEEEtran}
\bibliography{references_manos_TCOM}

\vfill

\end{document}